\documentclass[rnote]{aa} % for the research notes
\usepackage{graphicx}
\usepackage{txfonts}
\usepackage{epsfig}
\begin{document}
\title{Spectroscopic observations of the quiescent neutron star system 4U 2129+47 (= V1727 Cyg)}
      \author{M. S. Bothwell\inst{1,2}  
        \and 
        M. P. Torres\inst{3}
        \and          
        M. R. Garcia\inst{3}
        \and
        P. A. Charles\inst{4}
           }

\institute{Institite of Astronomy, University of Cambridge, Madingley Road, Cambridge, CB3 0HA
   \and
   School of Physics and Astronomy, Southampton University, Southampton SO17 1BJ   
\and
Harvard-Smithsonian Center for Astrophysics, Cambridge, MA 02138
\and
South African Astronomical Observatory, PO Box 9, Observatory, 7935, South Africa
   } 

\date{Received; accepted}

% \abstract{}{}{}{}{}
% 5 {} token are mandatory
  \abstract
{In quiescence, the proposed optical counterpart to the neutron
star low mass X-ray binary 4U 2129+47 (V1727 Cyg) shows a spectrum consistent with a late F-type subgiant and no radial
velocity variations on the 5.24 hour binary period. This could
imply that V1727 Cyg is a chance line of sight interloper. Radial
velocity measurements, however, showed evidence for a longer term
$\sim$40 km s$^{-1}$ shift, which suggested that 4U 2129+47 could be a
hierarchical triple system, with the F-type star in a wide
orbit about the inner low mass X-ray binary.}
{In order to confirm the long-term radial velocity shift reported in
Garcia et al. (1989) and its amplitude, we obtained spectroscopic
observations of V1727 Cyg during 1996 and 1998 with the William
Herschel Telescope using the ISIS spectrograph.}
{We determined radial velocities from the ISIS spectra by means of the
cross-correlation technique with a template spectrum.}
{The resulting radial velocities show variations with a maximum
amplitude of $\sim40$~km s$^{-1}$, confirming previous results and
supporting the F-type star as being the third body in a hierarchical triple system. The odds that this star could be an interloper are  $\sim 3 \times 10^{-6}$.}
  {}


\titlerunning{Spectroscopic  Observations of V1727 Cyg}

\maketitle

%
%________________________________________________________________

\section{Introduction}

4U 2129+47 was discovered as an active, but weak X-ray source in the
fourth UHURU survey (Forman et al. 1978), and its optical counterpart
was identified as V1727 Cyg (Thorstensen et al. 1979). V1727 Cyg was
found to be of $\sim$17th magnitude, and exhibited a large amplitude
($\Delta$B $\sim$1.5) photometric periodicity  of 5.24 hours. This was
taken to be indicative of the orbital period, with the photometric
modulation ascribed to X-ray heating of the companion (Thorstensen et
al. 1979). Analysis of the X-ray light curve led to the development of
a model where the system is viewed edge on and surrounded by an
accretion disc corona (White \& Holt 1982). Attempts to measure the
mass of the compact object produced confusing results, with radial
velocity studies implying a compact object mass of 0.6 $\pm$ 0.2
M$_{\odot}$ (Horne et al. 1986). This led to the conclusion that the
binary system was a cataclysmic variable. However, the discovery of an
X-ray burst confirmed the identity as a neutron star X-ray binary
(Garcia \& Grindlay 1987). Assuming a neutron star primary and a 5.24
hour period, the empirical relationship described by Robinson (1976)
between the companion mass and the binary period indicates a mass of $
0.59 \leq M (M_\odot) \leq 0.80$ and a radius of $0.59 \leq R
(R_\odot) \leq 0.68$ (Thorstensen et al. 1979). These parameters
suggest a K - M V spectral type for the companion star.

X-ray observations in 1983 failed to detect 4U 2129+47, and
contemporaneous optical observations showed that V1727 Cyg had dimmed
to $\sim$18.5th mag (Pietsch et al. 1986). This period of quiescence
was a good opportunity for detecting the companion star with a view to
constraining the orbital parameters of the system. The first such
study was conducted by Garcia et al. (1989). Spectra were taken during
four observing runs between June 1987 and October 1988 in order to
obtain more accurate measurements of the neutron star mass without the
strong influence of accretion disc emission. The results were
surprising: no variations on a 5.24 hour period were detected. From
the upper limits on the orbital radial velocity variations, the
neutron star's mass appeared to be $< 0.1 $ M$_{\odot}$ (Garcia et
al. 1989), which is obviously inconsistent with a neutron star
primary.  Moreover, the June 1987 run provided a mean velocity
significantly higher (by $\sim 40$~km s$^{-1}$) than observed in the
other nights. Another surprise 
was the identification of V1727 Cyg
as an F7-IV star - a stable 5.24 hour orbit about a neutron star
is smaller than the radius of an F7-IV star (by a factor of $\sim$
1.5). These results have led to the suggestion that 4U 2129+47 is a
hierarchical triple system. The observed F-type star is
postulated to be in a wide orbit about the centre of
mass of the close pair, while the inner companion is perhaps a K-type dwarf
(Garcia et al. 1989). The F7 star dominates the optical spectrum in
the current quiescent state, and so radial velocity shifts on a 5.24
hour period are not observed. Garcia et al. (1989) suggested that a 30 day outer period could account for the on-off cycles seen in 4U 2129+47. An outer period on this order would drive a periodic eccentricity in the inner binary on a 45 year timescale, consistent with the on and off states that have been observed since the 1930s. The radial velocity data previously obtained and newly reported herein are consistent with such a period.
Chevalier et
al. (1989)  have pointed out that even without the observed radial
velocity shift, the lack of large radial velocity variations would
lead to the postulation of a third stellar component.

An obvious alternative to the triple hypothesis is a chance line of
sight alignment. Ground based studies show that the on and off state
optical counterparts are coincident to $0.26''$ indicating that the
likelihood of a chance alignment is $ 10^{-3}$ (Thorstensen et al.
1988).  HST imaging shows that there is no nearby companion of
comparable magnitude within $0.04''$ (Deutsch et al. 1996).  Scaling
this radius to the ground based results further reduces the chance
alignment probability to $2 \times 10^{-5}$.

In this paper, we present spectra of V1727 Cyg taken in order to
analyse its systemic radial velocity over a time interval similar to
the long ($\sim 30$ day) period predicted for the late F-type
star if this is the outer component of a triple system (Garcia et
al. 1989). Previous radial velocity data were acquired during short
observing runs (over at most 2 nights), separated by periods of many
months. In order to test the proposed $\sim 30$ day period it was
necessary to conduct a study over a time interval of $\sim 1$ month,
with data points separated by about a week.

\section[]{Observations}

Optical spectra were acquired in service mode during 1996 and 1998
using the 4.2m William Herschel Telescope on La Palma equipped
with the dual-beam ISIS spectrograph. V1727 Cyg was observed with both
arms of the instrument.

The 1996 data set was acquired on August 5/12/19/25 UT. The
observations were made using the R1200R grating and TEK2 CCD in the
red arm and the R600B and TEK1 CCD in the blue arm, giving a
dispersion of 0.40~\AA~pixel$^{-1}$ and 0.78~\AA~pixel$^{-1}$,
respectively. The slit width used ranged from 1" .0 to 1".3. From
measurements of the arc-lamp and night sky emission lines, we find
that this yielded a spectral resolution from 1.0 to 1.2~\AA~(red arm)
and 1.4 to 2.0~\AA~(blue arm). The red and blue data cover the 
wavelength range $\lambda\lambda 6370 - 6760$ and $\lambda\lambda
4340-4910$. On each night, two to three 30 min spectra of V1727 Cyg
were taken, along with spectra of a template star, either BD+47
4219 or HD 222368 (both F7-IV stars). HD 222368 is a GCRV star with a
velocity of $5.0\pm0.9$ km s$^{-1}$. In total, 11 spectra of
V1727 Cyg were obtained. Calibration frames were also acquired, in
particular arc lamp spectra. However, the night of August 4 had no
flat fields taken. The data have been included in the analysis as the
signal to noise was high enough to mask the fixed pattern noise:
performing a cross correlation between an August 5 object spectra and
an identically extracted strip from a flat field produced no peak,
indicating that the fixed pattern noise is small. The night of August
12 (with 2 target spectra) was discarded, as the arc lamp spectra were
not rigorously taken close in time to the target spectrum, leading to
an uncertain wavelength calibration.

The 1998 data set consists of 6 nights of data: June 19/27, July
3/17/22 and August 2 UT. The observations were taken with the
R1200R grating and TEK2 CCD on the red arm and the R1200B grating and
EEV10 CCD on the blue arm of the spectrograph. The slit width used
ranged from  1".0 to 1".2. In the red arm, this provided a spectral
resolution from 0.8 to 1.5~\AA. The blue arm data was of poor
quality and were not used in the analysis. The red data covered the
 wavelength interval $\lambda\lambda 6310-6710$.  During each
night two to three spectra of V1727 Cyg with integration times
ranging from 15 to 30 min were taken, along with all associated
calibration frames. Spectra of BD+47 4219 were also acquired
each night (except on July 3). A total
of 15 spectra of V1727 Cyg were taken. The two spectra acquired
on June 27 were discarded due to their low quality. 
Furthermore, the night of July 16 was unsuitable for use due
to instrumental problems.

Both data sets were reduced with standard IRAF routines and the
spectra were extracted with the IRAF KPNOSLIT package.

\section[]{Analysis}

Velocities were computed using the Fourier cross-correlation technique
developed by Tonry \& Davis (1979) and implemented in the IRAF task
FXCOR. A sum of three HD 222368 spectra (taken on 1996 August 18 when
conditions were very good) was used as a template spectrum. Fig. 1
shows this composite spectrum from the red arm along with a sum of
four V1727 Cyg spectra. Prior to the cross-correlation, the target and
template spectra were re-sampled into a common logarithmic wavelength
scale and normalized by dividing with the result of fitting a
low-order spline to the continuum.  Correlation was performed without
H$\alpha$ in the red, to ensure that any possible residual emission
from the accretion disc did not interfere with the measured velocity
of V1727 Cyg. In the blue, H$\beta$ was included, as (besides the
G-band) the spectra lack well defined features in the available
region, and without H$\beta$ correlation fits were poor.

Fig. 2 and Table 1 show the radial velocity of V1727 Cyg from
the 1996 and 1998 data sets. The radial velocities shown in
Fig. 2 are weighted averages of radial velocities obtained from the
red and (when available) blue spectra. All velocities are
heliocentric. Table 1 also shows the average radial velocities
obtained as derived using only the blue or ned spectra. As in Garcia
et al. (1989) the errors on the individual velocities produced by
FXCOR have been computed as $\sigma =C/(1+r)$, where $C$ is a constant
determined from the observed scatter in the velocity of comparison
stars, and $r$ is the correlation coefficient (see Tonry and Davis
1979). Gaussian statistics were used throughout when calculating mean
values. Checking the wavelength of night sky emission lines served as
a zero point calibration for the data: these lines are seen only at
wavelengths longer than $\sim5000$~\AA, so this method could not be
used for the blue frames. The average night sky line velocity is
approximately -0.4 km s$^{-1}$, which is consistent with zero given
the errors.  The wavelength of each night sky line was found by
finding the centroid of a Gaussian function fitted to each line. Both
the larger error bar for the V1727 Cyg radial velocity and the lack of
sky line measurements on 1998 July 3 are due to the fact that the
spectra were obtained during dawn. As noted in section 2, no radial
velocity templates were observed during this night.

\section{Discussion}

Our data do not show the expected short term, $\sim 300$~km s$^{-1}$
semi-amplitude radial velocity variations indicative of a neutron star
primary. This is consistent with all previous studies of V1727 Cyg in
quiescence (see e.g. Garcia et al. 1989; Garcia 1989; Chevalier et
al. 1989; Cowley \& Schmidtke 1990). Furthermore, the longer term
velocity variations found by Garcia et al. 1989 are confirmed in this
new dataset.  The 40 km s$^{-1}$ amplitude of the variation is similar
to that found earlier, and the variation is detected between 7
and 10~$\sigma$ level.

If we accept that the late F-type subgiant is indeed
orbiting an unseen companion (or companions) we can ask what are the
odds that it could still be a line of sight interloper, rather than
physically associated with 4U 2129+47? The fraction of multiple systems
in the Galaxy is an active topic of study, with estimates ranging from
10\% to 80\% (Lada 2006). It is now clear that the fraction depends
upon mass, with more massive stars more likely to be members of
multiple systems. Also, the majority of the systems are widely
detached and show only small (few km s$^{-1}$) radial velocity
variations.  We used the study of Duquennoy \& Mayor (1991) to
estimate the fraction of F7 stars that would show velocity variations
of at least 40 km s$^{-1}$ (corresponding to a radial velocity
semi-amplitude $K$ of 20 km s$^{-1}$).  This study measured orbital
velocity variations of a sample of F7 to G8 stars in order to
determine orbital periods and multiplicity frequency.  The mean period
in the sample was $log(P) = 4.8$ with a roughly Gaussian dispersion of
$log(P) = 2.3$ with $P$ in days.  From Table 2 of Duquennoy \& Mayor
(1991) we estimate that $K = 20$~km s$^{-1}$ corresponds approximately
to $P = 100$~days. The fraction of systems with periods less than
100~days, and therefore likely to show velocity variation of 40 km
s$^{-1}$ or more, is approximately 1 in 7.  Therefore the odds that
V1727 Cyg could be an interloper are reduced from $2 \times 10^{-5}$
to approximately $3 \times 10^{-6}$.

Recently, Bozzo et al. (2007) have presented time analysis of two
XMM-Newton observations of 4U 2129+47 in quiescence obtained $\sim
22$~days apart, finding evidence for a delay of $\sim 200$~s for the
mid-eclipse times measured from the X-ray observations. This delay
can be explained as due to the orbital motion of the compact 4U 2129+47
binary around the center of mass of a triple system. In light of these
new observations, coupled with the radial velocity measurements
presented in this paper, we conclude that the triple hypothesis best
explains all features of 4U 2129+47, and that the late F-type subgiant
is the outer component of the triple system.

\section{Conclusions}

The X-ray source 4U 2129+47 has been a candidate hierarchical triple
for some time. The optical counterpart, V1727 Cyg, has features
inconsistent with those expected for a close binary companion, and
previously displayed radial velocity variation over several weeks. The
new spectroscopic measurements presented here confirm this long term
shift of $\sim$ 40 km s$^{-1}$, which strongly suggests that 4U 2129+47
is indeed a triple.  However, the outer period is yet to be observed
directly.  Doing so would allow for more sophisticated modeling of the
system. It may also help understand the evolutionary history of
4U 2129+47, which as of now is somewhat elusive.

\begin{acknowledgements}

This paper makes use of data obtained from the Isaac Newton Group
Archive which is maintained as part of the CASU Astronomical Data
Centre at the Institute of Astronomy, Cambridge.  We acknowledge NASA/
Chandra Data Center Contract NAS8-03060, which provides partial
support for MRG.

\end{acknowledgements}

%______________________________________________________________

\

\begin{table}
\caption{Heliocentric radial velocities in km s$^{-1}$) for V1727 Cyg in quiescence.}
\label{}
\begin{center}
\begin{tabular}{cccc}
\hline\hline
Date & Radial Velocity    & Radial Velocity    & Radial Velocity      \\
(UT) & (blue and Red arm) & (blue arm)         & (red arm)\\
\hline 
1996 Aug 5.0  & $-48.4 \pm 5.8$  & $-48.7 \pm 7.8$  & $-48.3 \pm 2.7$\\
1996 Aug 19.2 & $-45.4 \pm 5.1$  & $-48.3 \pm 6.0$  & $-39.3 \pm 3.4$\\
1996 Aug 25.1 & $-8.0  \pm 3.9$  & $-3.1  \pm 4.2$  &$-14.4 \pm 4.0$\\
1998 Jun 19.2 & $-40.3 \pm 3.6$  & ---              & $-40.3 \pm 3.6$\\
1998 Jul 3.2  & $-5  \pm 20$     & ---              &$-5 \pm 20$ \\
1998 Jul 22.1 & $-17.2 \pm 5.2$  & ---              &$-17.2 \pm 5.2$\\
1998 Aug 2.1  & $-22 \pm 11$   & ---              &$-22 \pm 11$\\ 
\hline
\end{tabular}
\end{center}
\end{table}

\begin{figure}[]
\label{fig:comp}
\centerline{\includegraphics[scale=0.32,angle=0]{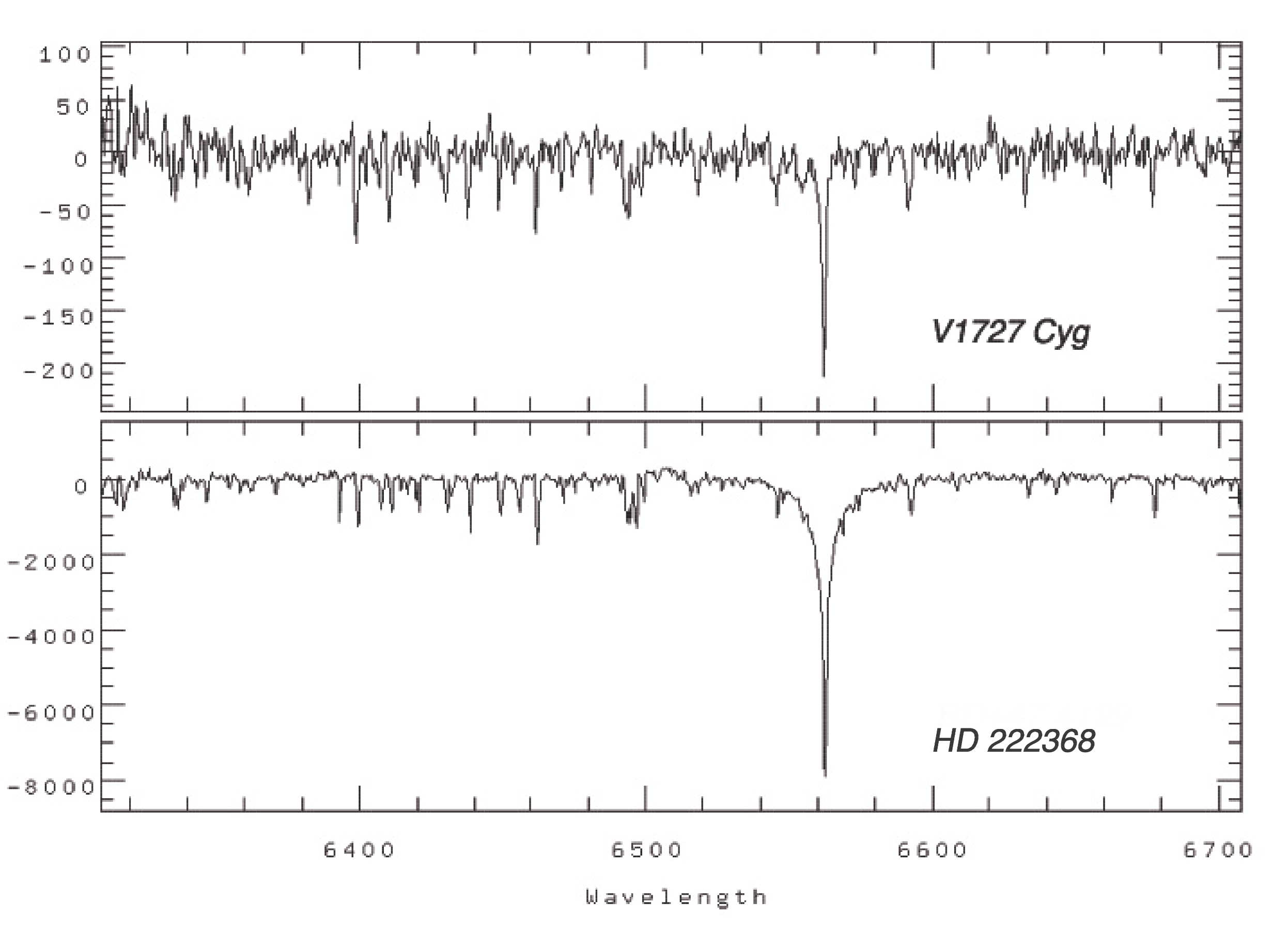}}
\caption{Summed spectra of V1727 Cyg compared to that of the template
spectrum used, HD222368.}
\end{figure}

\begin{figure}[]
\label{fig2:comp}
\centerline{\includegraphics[scale=0.32,angle=0]{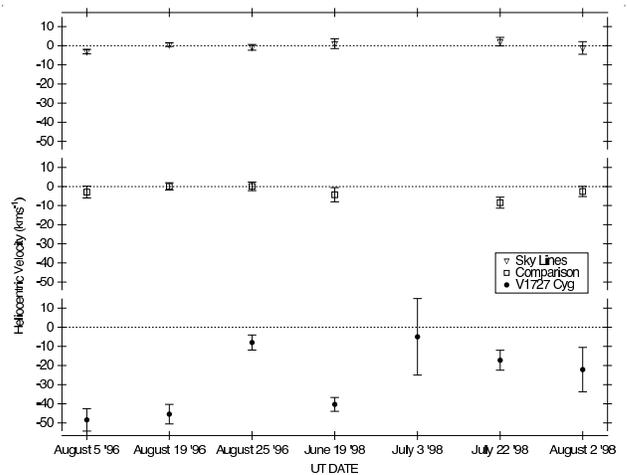}}
\caption{New radial velocity data for V1727 Cyg (bottom) from 1996 and 1998
(see Table 1). Also plotted are sky line velocities as a zero point
check (top plot). These velocities were calculated from the difference between the
measured and published (Osterbrock et al. 1996) wavelengths of sky
emission lines. Comparison star velocities are shown in the middle
plot.}
\end{figure}


\begin{thebibliography}{}



\bibitem[2007]{Bozzo2007}Bozzo, E., Falanga, M., Papitto, A., et al.  
   2007, A\&A, 476, 301


\bibitem[1989]{Chevalier1989} Chevalier, C., Ilovaisky, S.~A., Motch, C., et al. 
   1989, A\&A, 217, 108

\bibitem[1990]{Cowley1990} Cowley A. P., \& Schmidtke P. C. 
   1990, AJ, 99, 678

\bibitem[1996]{Deutsch1996} Deutsch, E. W, Margon, B., Wachter, S., et al. 
   1996, ApJ, 471, 979

\bibitem[1991]{Dunquennoy1991} Duquennoy, A. \& Mayor, M. 
   1991, A\&A, 248, 485 

\bibitem[1978]{Forman1978} Forman, W., Jones, C., Cominsky, L., et al. 
   1978, ApJSS, 38, 357

\bibitem[1987]{Garcia1987} Garcia, M. R., \& Grindlay, J. E. 
   1987, ApJ, 313, L59

\bibitem[1989]{} Garcia, M. R., Bailyn, C. D., Grindlay J. E., et al. 
   1989, ApJ, 341, L75

\bibitem[1989]{} Garcia, M. R. 1989, Proc. 23rd ESLAB Symp., ESA SP-296, 151


\bibitem[1986]{Horne1986} Horne, K., Verbunt, F., Schneider, D. P.
   1986, MNRAS, 218, 63

\bibitem[2006]{Lada2006} Lada, C. J. 
   2006, ApJ, 640. L63


\bibitem[1996]{Osterbrock1996} Osterbrock, D. E., Fulbright, J. P., Martel, A. R, et al. 
   1996, PASP, 108, 277   

\bibitem[1986]{Pietsch1986} Pietsch, W., Steinle, H., Gottwald, M., et al. 
   1986, AJ, 157, 23


\bibitem[1976]{Robinson1976} Robinson, E. L. 
   1976, ApJ, 203, 485

\bibitem[1979]{Thorstensen1979} Thorstensen, J. Charles, P., Bowyer, S. et al. 
   1979, ApJ, 233, L57

\bibitem[1988]{Thorstensen1988} Thorstensen, J. R., Brownsberger, K. R., Mook, D. E., Berg, C.; Bramson, J., et al. 
   1988, ApJ, 334, 430

\bibitem[1979]{Tonry1979} Tonry, J., \& Davis, M. 
   1979, AJ, 84, 1511

\bibitem[1982]{White1982} White, N., \& Holt, S. 
   1982, ApJ, 257, 318

\end{thebibliography}
\end{document}